\documentstyle[12pt,epsf]{article}
\begin{document}

\begin{titlepage}
\begin{flushright}
IPNO/TH 96-13\\
\end{flushright}
\vfill
\centerline{\large\bf HEAVY BARYON CHIRAL PERTURBATION THEORY}
\vspace*{0.1cm}
\centerline{\large\bf AND THE SPIN 3/2 DELTA RESONANCES
$\footnote{Talk given at 7th International Conference on the Structure of 
Baryons, Santa Fe, NM, 3-7 Oct 1995.}$
}
\vfill
\centerline{JOACHIM KAMBOR}
\vspace*{0.5cm}
\centerline{Division de Physique Th\'eorique 
$\footnote{Unit\'e de Recherche des Universit\'es Paris XI et Paris VI, 
associ\'e au CNRS.}$, 
Institut de Physique Nucl\'eaire} 
\centerline{F-91406 Orsay Cedex, France}
\vfill
\begin{abstract}
Heavy baryon chiral perturbation theory is briefly reviewed, paying particular
attention to the role of the spin 3/2 delta resonances. The concept of 
resonance saturation for the baryonic sector is critically discussed. 
Starting from a relativistic formulation of the pion-nucleon-delta system,
the heavy baryon chiral lagrangian including spin 3/2 resonances is constructed
by means of a 1/m-expansion. The effective theory obtained admits a systematic 
expansion in terms of soft momenta, the pion mass $M_\pi$ and the 
delta-nucleon mass difference $\Delta$.
\end{abstract}
\vfill
\end{titlepage}
\newpage

\section{Introduction}

Chiral symmetry restricts severely the interactions of pions, nucleons and 
photons \cite{CA}. The consequences are most conveniently summarized by 
the use of
an effective field theory, valid in the low energy regime. The simultaneous 
expansion in small momenta and light quark masses is known as Chiral 
Perturbation Theory (ChPT) \cite{Wei79,GL85a,GSS88}. Any parameter free 
prediction of ChPT, to a given order in the low energy expansion, is also
a prediction of low energy QCD. However, since the theory is nonrenormalizable,
there appear new low energy coupling constants, which parametrize the dynamics
of the underlying theory and are not fixed by the symmetry requirement. In 
practice, in order to make accurate predictions, some of these coupling 
constants must be fixed, either from experiments or from further theoretical 
considerations, such as resonance saturation. Our ignorance about the value 
of these so called counter term coupling constants constitutes the main 
limitation of the method.

Extending ChPT to the baryonic sector, there are two issues which need to be 
discussed further. 
First, the mass of the nucleon is large and does not vanish in the chiral 
limit. This spoils a constistent chiral power counting. The problem can 
be overcome by treating the nucleon as a very heavy source \cite{JM91}, or, 
equivalently, by performing a systematic $1/m$-expansion \cite{BKKM92}, 
$m$ being the nucleon mass. The 
expansion is known as Haevy Baryon Chiral Perturbation Theory (HBChPT).
The second issue I will address is the 
role the spin $3/2$ delta resonances play in the pion-nucleon system. 
The mass difference $\Delta=m_\Delta-m$
is not large compared to the typical low energy scale $M_\pi$. Moreover,
the singularities 
associated with the delta are strongly coupled. 
Two approaches can be found in the literature . The first of these 
takes into account the effect of the delta (and of other 
resonances) only through contributions to the coupling constants of higher 
order operators in the chiral expansion \cite{BKMrev95}. 
This approach is in particular well suited to derive low energy theorems. 
The concern, however, is that, in the physical 
world of nonvanishing quark masses, the perturbation series might converge 
slowly due to large coupling constants driven by small denominators---{\it
i.e.} by terms proportional to $1/\Delta$. The second approach to HBChPT 
includes the delta degrees of 
freedom explicitly \cite{JM91,JMDob}. In addition to solving 
the problems mentioned above, this technique has 
the advantage that the range of applicability is in principle 
extended into the delta-region. However, care must be taken in order to 
fully implement the requirements of relativistic quantum field theory. 

The plan of the talk is as follows. In section 2, the formulation of heavy 
baryon chiral perturbation theory is briefly reviewed. In section 3, the role
of the delta resonances is critically examined. Particular emphasis is put
on conceptual problems concerning 
resonance saturation in the presence of the delta resonances. A systematic 
$1/m$-expansion of the $\pi N \Delta$-system, based on recent work in 
collaboration with Hemmert and Holstein \cite{HHK96a,HHK96b}, is presented. 
Two examples are given which illustrate the questions involved. 
 
\section{Heavy Baryon Chiral Perturbation Theory}

We start with the relativistic formulation of ChPT for the $\pi N$-system
and write the effective lagrangian as a string of terms \cite{GSS88}
\begin{equation}
{\cal L}_{\pi N}={\cal L}_{\pi N}^{(1)}+{\cal L}_{\pi N}^{(2)}+ \ldots
\end{equation}
where the superscript denotes the number of derivatives. The first term in this
expansion reads 
\begin{equation}
{\cal L}_{\pi N}^{(1)}=\bar N \left( i\not\!\!{D}-\dot{m}+i{\dot{g}_A \over 2} 
\not\!{u} \gamma_5 \right) N
\label{LpiN1}
\end{equation}
where $N$ is the nucleon field, $\dot{m}$ and $\dot{g}_A$ are the nucleon 
mass and axial-vector coupling constant in the chiral limit, respectively.

In the chiral limit, $\dot{m}\not=0$. As a consequence, the covariant derivative
on $N$ counts as order one ($p$ denotes a generic soft momentum)
\begin{equation}
{D}_\mu N= O(1) \qquad{\rm but}\qquad
(i\not\!\!{D}-\dot{m}) N = O(p).
\end{equation}
Therefore, the loop-expansion is not equivalent to a low-energy expansion
\cite{GSS88}, as in the Goldstone boson sector \cite{Wei79}.
The problem can be overcome by going to the extreme nonrelativistic limit
\cite{JM91,JMDob}.
The idea is to move the $\dot{m}$-dependence from the propagator to the 
vertices. One way to achieve this is to decompose the nucleon four-momentum
according to
\begin{equation}
p_\mu=\dot{m} v_\mu +k_\mu, \qquad v^2=1,
\end{equation}
where $k_\mu$ is a soft residual momentum. The upper component of the 
nucleon field in its restframe, $H_v$, defined by
($P_v^\pm=(1\pm \not\!{v})/2$)
\begin{equation}
N = \exp\{-i\dot{m} (v\cdot x)\} (H_v+h_v), \qquad
P_v^+ H_v = H_v, \quad
P_v^- h_v = h_v
\end{equation} 
obeys $i\not\!{\partial} H_v= \not\!{k} H_v$.
The derivative on the field $H_v$ is thus of order $p$. Writing the effective 
lagrangian in terms of fields 
$H_v$ only ensures a consistent low energy expansion. The leading order term 
becomes
\begin{equation}
{\cal L}_{\pi N}^{(1)}=\bar H_v \left( i (v\cdot D)
+\dot{g}_A (S\cdot u) \right) H_v
\end{equation}
where $S_\mu$ denotes the Pauli-Lubanski spin vector. The propagator reads
\begin{equation}
S(\omega)={i \over v\cdot k +i\epsilon},
\end{equation}
with $\omega=v\cdot k$. The low energy expansion of the $\pi N$-system 
so obtained is a simultaneous expansion in 
\begin{equation}
{p\over 4\pi F_\pi} \qquad {\rm and} \qquad {q\over \dot{m}}\ . 
\end{equation}
This is most easily seen by using a path integral formulation where the heavy 
degrees of freedom, $h_v$, are systematically integrated out \cite{BKKM92}. 
The approach also shows, that some of the low energy coupling constants must 
have some specified values, otherwise the theory would not be Lorentz 
invariant \cite{EM96}. The last observation can alternatively be derived 
by employing reparametrization invariance \cite{LM92}. Renormalization at 
the one-loop level (up to order $p^3$) is thoroughly discussed in ref. 
\cite{E94}. 

\section{Inclusion of the delta resonances: systematic 1/m-expansion}

The spin 3/2 resonances $\Delta$(1232) play a special role in the 
$\pi N$-system. Whenever delta exchange contributes to a given obsevable 
(not forbidden by quantum numbers) the effect is large. The mass difference
\begin{equation}
\Delta\equiv m_\Delta-m_N \approx 300 {\rm MeV}
\end{equation}
is small and of the same order as a typical low energy scale, e.g. $M_\pi$. 
Therefore, the delta resonances cannot be treated as heavy compared to the 
nucleon. The concept of resonance saturation is in this case at least 
questionable and it's applicability warrants further discussion.

In order to formulate the problem more precisely, consider the magnetic
polarizability of the nucleon in HBChPT. It has a low energy expansion of 
the form (modulo logarithms of $M_\pi$) \cite{BKKM92}
\begin{equation}
\beta={{\rm const.} \over M_\pi} \left\{ 1+c_1 {M_\pi\over\Lambda}
+c_2 {M_\pi^2 \over \Lambda^2}+\ldots \right\}
\label{betaexp}
\end{equation}
where $\lambda\in \left\{ 4\pi F_\pi, m_N\right\}$ is a heavy scale and 
the $c_i$ are dimensionless constants. This expansion is well suited to 
derive low-energy theorems (LET)\cite{EMLet95}, valid in the chiral limit, 
i.e.
\begin{equation}
\lim_{M_\pi \rightarrow 0} M_\pi \beta= {\rm const.} 
\end{equation}
In the physical world of finite quark masses, however, the series 
(\ref{betaexp}) might converge slowly, due to large coefficients $c_i$.
Consider the effect of delta exchange on $\beta$, as shown in Fig. 1.
If we shrink the delta propagator to a point, the constants $c_i$ will
pick contributions of the form
\begin{equation}
c_i\sim \left( {m_\rho \over \Delta } \right)^i,
\end{equation}
\begin{figure}
\begin{center}
\leavevmode
\hbox{%
\epsfxsize=10.0truecm
\epsfbox{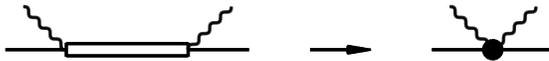} }
\end{center}
\caption{$\Delta$-resonance exchange contribution to nucleon Compton 
scattering. Single,
double and wavy lines denote nucleons, delta and photons, respectively. The 
shaded dot denotes a local counterterm.}
\end{figure}
where I have inserted $m_\rho$ as a typical hadronic scale. In such a 
circumstance, the individual terms in the bracket in (\ref{betaexp}) are
all of order one and it will be necessary to resum the series. The whole 
point is that the scale $\Delta$ appears in the denominator, not $m_\Delta$. 
If we consider the limit $m_\Delta\rightarrow \infty$, we also have to let
$m_N\rightarrow \infty$, with $\Delta$ fixed.

The contribution of the delta to $\pi N$-processes via exchange graphs does 
not exhaust the low energy manifestations of the delta's. 
\footnote{The counterterms of ${\cal L}_{\pi N}$ receive also contributions
from meson resonance exchange in the t-channel, see \cite{BKMrev95}, sect. 3.4.}
There are further nearby singularities generated by loop-graphs, as shown 
in Fig. 2 a). In the conventional formulation of HBChPT, such graphs should be 
accounted for in the relativistic framework \cite{BM96}. However, as there 
is no consistent powercounting in this framework, one might consider even 
higher loop graphs, as shown in Fig. 2 b). It is presently not clear how in 
this instance
resonance saturation of counterterms can be formulated in a systematic manner.
\begin{figure}
\begin{center}
\leavevmode
\hbox{%
\epsfxsize=10.0truecm
\epsfbox{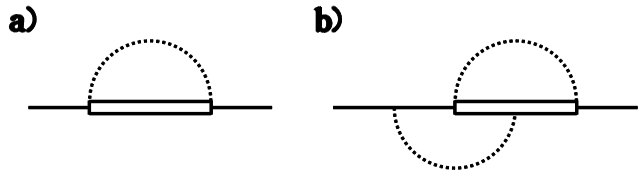} }
\end{center}
\caption{Contributions to the nucleon selfenergy with deltas in the loop: 
a) leading one-loop graph, b) an example of higher loop-graphs.}
\end{figure}

The conceptual problems just described are related to the fact that the
delta resonances (and also the nucleon!) contain both, heavy and light degrees
of freedom. The light components of the delta resonances decouple only in 
the strict chiral limit \cite{JMDob}. Jenkins and Manohar have included 
the Baryon
decuplet as dynamical degree of freedom in the formulation of HBChPT right from
the beginning \cite{JM91}. Here I will describe recent work in collaboration 
with Hemmert and Holstein \cite{HHK96a,HHK96b}, where we show explicitly, 
by means of a systematic $1/m$-expansion, how the framework of \cite{JM91} can 
be obtained from the relativistic formulation of the $\pi N\Delta$ 
interactions. 

Consider the lagrangian for a relativistic spin 3/2 field $\Psi_\mu$ coupled 
in a chirally invariant manner to the Goldstone bosons
\footnote{For the sake of simplicity, we suppress isospin indices throughout. 
Also, we consider exclusively chiral SU(2) in this article.}
\begin{eqnarray}
{\cal L}_{3/2}&=&\bar{\psi}^\mu\Lambda_{\mu\nu}\psi^\nu\nonumber\\
\Lambda_{\mu \nu} &=& - \; [ ( i \not\!\!{D} - 
                              m_{\Delta} ) g_{\mu \nu} - \frac{1}{4} 
                              \gamma_{\mu} \gamma^{\lambda} ( i \not\!\!{D} - 
                              m_{\Delta} ) \gamma_{\lambda} \gamma_{\nu} 
                              \nonumber \\
                  && + \frac{g_{1}}{2} g_{\mu \nu} \not\!{u} 
                              \gamma_{5} + \frac{g_{2}}{2} ( \gamma_{\mu} 
                              u_{\nu} + u_{\mu} \gamma_{\nu} ) \gamma_{5} 
                      + \frac{g_{3}}{2} \gamma_{\mu} \not\!{u} 
                              \gamma_{5} \gamma_{\nu} ]. 
\label{eq:Lambda}
\end{eqnarray}
We have factored out the dependence on the unphysical free parameter A using
$\psi_\mu(x)=(g_{\mu\nu}+ A/2 \gamma_\mu \gamma_\nu)\Psi^\nu(x)$ \cite{Pas94}. 
The pion fields are contained in 
$u= \exp \left(i\vec{\tau} \cdot \vec{\pi}/F_\pi \right)$
and $D_\mu\psi_\nu$ denotes the covariant derivative 
on the spin 3/2 field.
The first two pieces in (\ref{eq:Lambda}) are the kinetic and
mass terms of a free Rarita-Schwinger Spinor \cite{BDM89}. The remaining terms
constitute the most general chiral invariant couplings to pions. 

The next step consists of identifying the light and heavy degrees of freedom
of the spin 3/2 fields, respectively. The procedure is analogous to the case 
of spin 1/2 fields \cite{BKKM92}, but technically more complicated 
due to the off-shell spin 1/2 degrees of freedom of the Rarita-Schwinger
field. In order to get rid of the mass dependence in (\ref{eq:Lambda}) 
we introduce the spin 3/2 projection operator for fields with {\it 
fixed velocity} $v_\mu$
\begin{equation}
P^{3/2}_{(33)\mu \nu} =  g_{\mu \nu} - \frac{1}{3} \gamma_{\mu} \gamma_{
                               \nu} - \frac{1}{3} \left( \not\!{v} \gamma_{\mu} 
                               v_{\nu} + v_{\mu} \gamma_{\nu}\not\!{v} \right) 
\label{eq:project}
\end{equation}
and identify the light degrees of freedom via
\begin{equation}
T_{\mu,v} (x) \equiv  P_{v}^{+} \; P^{3/2}_{(33)\mu\nu} \psi^{\nu}(x) 
               \; \exp (i \dot{m} v \cdot x) . 
\label{eq:T}
\end{equation}
These fields coincide with the decuplet fields used in \cite{JM91} and 
satisfy $v_\mu T^\mu_v=\gamma_\mu T^\mu_v=0$.
The remaining components, denoted collectively by $G_{\mu, v}$, are heavy 
and will be integrated out.
However, the effects of these degrees of freedom are 
included as they will give rise to $1/m$-corrections.

We now perform a systematic $1/m$-expansion, in analogy to the heavy 
nucleon formalism \cite{BKKM92}.  
We write the most general lagrangian as (suppressing all indices)
\begin{equation}
{\cal L}={\cal L}_{N} + {\cal L}_{\Delta} + {\cal L}_{\Delta N}
\end{equation}
with
\begin{eqnarray}
{\cal L}_{N} &=& \bar{H} {\cal A}_{N} H + ( \bar{h} {\cal B}_{N} H + h.c. ) + 
                  \bar{h} {\cal C}_{N} h \nonumber \\
{\cal L}_{\Delta N} &=& \bar{T} {\cal A}_{\Delta N} H + \bar{G} 
{\cal B}_{\Delta N} H + 
                         \bar{h} {\cal B}_{N \Delta} T + \bar G 
{\cal C}_{\Delta N} h +
                         h.c. \nonumber\\
{\cal L}_{\Delta}&=&\bar{T} {\cal A}_{\Delta} T + \bar{G} 
              {\cal B}_{\Delta} T + \bar{T} \gamma_{0} {\cal 
              B}_{\Delta}^{\dagger} \gamma_{0} G + \bar{G} {\cal C}_{\Delta} G.
\label{Lgeneral}
\end{eqnarray}

The matrices ${\cal A}_{N}$, ${\cal B}_{N}$, ...,${\cal C}_{\Delta}$ 
in \ref{Lgeneral} admit an expansion of the form
\begin{equation}
{\cal A}_\Delta = {\cal A}_\Delta^{(1)}+{\cal A}_\Delta^{(2)}+ ... ,
\end{equation}
where ${\cal A}_\Delta^{(n)}$ is of order $\epsilon^n$. Here we denote 
by $\epsilon$ small quantities of order $p$, like $M_\pi$ or soft momenta, 
as well as the mass difference $\Delta\equiv m_\Delta-m$. In the physical world,
$\Delta$ and $M_\pi$ are of comparable magnitude. We therefore adhere to a 
simultaneous expansion in both quantities.  It is only through this 
(''small scale``) expansion that we obtain a systematic expansion of 
the $\pi N \Delta$-system. 

To make this more explicit, consider the leading order contribution to 
${\cal A}_\Delta$
\begin{equation}
{\cal A}_{\Delta, \mu\nu}^{(1)} =
- \left[ i (v \cdot D)-\Delta + g_{1} (S \cdot u) \right] \; g_{\mu \nu}.
\label{llead}
\end{equation}
The heavy baryon propagator of the delta is thus 
$-i P^{3/2}_{(33)\mu\nu}/((v\cdot k)-\Delta)$ 
and hence counts as order $\epsilon^{-1}$ in our
expansion. Explicit expressions for the expansions of
${\cal B}_\Delta$, ${\cal C}_\Delta$ etc. can be found in \cite{HHK96b}. 
Note that matrices ${\cal C}_\Delta$ and ${\cal C}_N$ start at order 
$\epsilon^0$.

The final step is again in analogy to the case where only nucleons were 
considered. Shifting variables and completing the square we obtain
the effective action
\begin{equation}
S_{\rm eff}= \int d^4x \left\{ \bar T \tilde {\cal A}_{\Delta} T
+\bar H \tilde {\cal A}_{N} H
+\left[ \bar T \tilde {\cal A}_{\Delta N} H + h.c.\right] \right\}
\label{Seff}
\end{equation}
with (I keep only leading order terms in $1/m$ here, for the sake of simplicity)
\begin{eqnarray}
\tilde {\cal A}_\Delta &=& {\cal A}_\Delta 
-\gamma_0  {\cal B}_{N \Delta}^\dagger \gamma_0 {\cal C}_N^{-1} 
{\cal B}_{N \Delta}
-\gamma_0 {\cal B}_\Delta^\dagger \gamma_0 {\cal C}_\Delta^{-1} {\cal B}_\Delta
\nonumber\\
\tilde {\cal A}_N &=& {\cal A}_N 
-\gamma_0  {\cal B}_{N}^\dagger \gamma_0 {\cal C}_N^{-1} 
{\cal B}_{N}
-\gamma_0 {\cal B}_{\Delta N}^\dagger \gamma_0 {\cal C}_\Delta^{-1} 
{\cal B}_{\Delta N}
\nonumber \\
\tilde {\cal A}_{\Delta N} &=& {\cal A}_{\Delta N} 
-\gamma_0  {\cal B}_{N \Delta}^\dagger \gamma_0  {\cal C}_N^{-1} 
{\cal B}_{N} 
-\gamma_0 {\cal B}_\Delta^\dagger \gamma_0 {\cal C}_\Delta^{-1} 
{\cal B}_{\Delta N} 
\label{Atilde}
\end{eqnarray}
The new terms in proportion to ${\cal C}_\Delta^{-1}$ and ${\cal C}_N^{-1}$
are given entirely in terms of coupling constants of the lagrangian for 
relativistic fields. This guarantees reparameterization
invariance \cite{LM92} and Lorentz invariance \cite{EM96}. Also, these terms 
are $1/m$ suppressed. The effects of the heavy degrees of freedom 
(both spin 3/2 and 1/2) thus show up only at order $\epsilon^2$. Note also that 
the effective $NN$-, $N\Delta$- and $\Delta\Delta$-interactions all contain
contributions from both heavy $N$- and $\Delta$-exchange respectively.  

In the above formalism, it is understood that one has to include also the most 
general counterterm lagrangian consistent with chiral symmetry, Lorentz 
invariance, and the discrete symmetries P and C, in relativistic formulation.
The construction yields then 
automatically the contributions to matrices ${\cal A}$, 
${\cal B}$, and ${\cal C}$. 
In order to calculate a given process to order 
$\epsilon^n$, it then suffices to construct matrices ${\cal A}$ to the same 
order, $\epsilon^n$,
${\cal B}$ to order $\epsilon^{n-1}$, and ${\cal C}$ to order $\epsilon^{n-2}$. 
Finally one has to add all loop-graphs contributing at the order one is
working. The relevant diagrams can be found by straightforward power 
counting in $\epsilon$.

\section{Applications}

In this section I discuss two simple applications in order to illustrate some
important features of the formalism. Consider first the nucleon mass to 
order $\epsilon^3$ in the low scale expansion. Starting from 
(\ref{Seff}) it is easily shown that, compared to the order $q^3$ calculation
in conventional HBChPT,  the only new contribution to the selfenergy is due
to the diagram shown in Fig. 2 a), where the $\Delta$-propagator denotes the 
propagation of the light degrees of freedom $T_{\mu,v}$. The result has been 
given already in 
\cite{BKM93}, the interpretation, however, is completely different. Thus
\begin{equation}
m_N=\dot{m}^r-4 c_1^r M_\pi^2-{3 \dot{g}_A^2 M_\pi^3 \over 32 \pi F_\pi^2}
+{g_{\pi N\Delta}^2 \over 12 \pi^2 F_\pi^2} R(\Delta, M_\pi)
\label{mN}
\end{equation}
where 
\begin{eqnarray}
R(\Delta, M)&=&-4(\Delta^2-M^2)^{3/2} \ln\left( {\Delta \over M}
+\sqrt{ {\Delta^2 \over M^2}-1 } \right) \nonumber\\
&&+4 \Delta^3 \ln {2\Delta \over M} 
-\Delta M^2 \left( 1+6 \ln {2 \Delta\over M}\right) .
\label{defR}
\end{eqnarray}
In (\ref{mN}) I have absorbed terms in proportion to $\Delta^3$ and 
$\Delta M_\pi^2$ in the counterterms $\dot{m}^r$ and $c_1^r$ respectively. 
These effects are unobservable as long as only the nucleon mass shift is
considered. Note, however, that the conterterm $c_1^r$ appaers also in 
the pion-nucleon sigma term. The function $R$ has the small $M$ expansion
\begin{equation}
R(\Delta, M)=-{9 \over 8} {M^4\over \Delta} 
\left( 1+{4\over 3}\ln {2\Delta\over M} \right) 
+O\left({M^6\over \Delta^3}\right),
\label{Rexp}
\end{equation}
which shows explicitly the decoupling of the delta resonances in the chiral 
limit. Consider now the function $R$ for physical masses 
$M_\pi=140\ {\rm MeV}$ and $\Delta\approx 2 M_\pi$. Each individual term in 
(\ref{defR}) is of order $\epsilon^3$, and $R$ suffers from large cancellations.
Numerically we find 
$R(\Delta, M_\pi)=-0.0044\ {\rm GeV}^3$. This corresponds to a nucleon mass 
shift of $-10\ {\rm MeV}$, comparable to the $-15\ {\rm MeV}$ shift coming 
from the 
term in proportion to $M_\pi^3$. Finally, we can study the convergence of the 
chiral expansion by taking into account only the leading term of order 
$M_\pi^4/\Delta$ in (\ref{Rexp}). This approximation yields again 
$R=-0.0044 \ {\rm GeV}^3$, i.e. the higher order terms in R can safely be 
neglected. We conclude that, for the nucleon mass, it is sufficient to work
to order $p^4$ in the chiral expansion. Note, however, that the 
low scale expansion presented here treats effects of comparable size,
$M_\pi^3$ and $M_\pi^4/\Delta$, at the same order of the expansion, 
viz. $\epsilon^3$. Not surprisingly, this in general improves the convergence 
of the perturbation series. 

As a second example I consider neutral pion photoproduction at threshold. 
To order $q^3$, the electric dipole amplitude obeys the LET
\cite{BGKM91,BKKM92}
\begin{equation}
E_{0+}(s_{\rm thr})=-{e g_{\pi N} \over 8\pi m} \mu \left\{
1-\left[ {1\over 2}(3+\kappa_p)+\left({m\over 4 F_\pi}\right)^2\right]\mu
+{\cal O}(\mu^2) \right\}
\label{LET}
\end{equation}
with $\mu=M_\pi/m$ as well 
as $\kappa_p$ the anomalous magnetic moment of the proton and $g_{\pi N}$
the strong pion-nucleon coupling constant. Recently an order $p^4$ 
calculation has been presented \cite{BKM94,Mei95} which reconciles the 
theoretical prediction with experiment \cite{Fuchs95}. However, each term in 
the expansion
of $\mu$ is large with alternating sign, thus making the convergence of this 
particular quantity very slow. 

Including now the delta degrees of freedom and working to order $\epsilon^3$,
there are no loop graphs with intermediate deltas contributing
at threshold. However, there is a new contribution via the pole graph of 
Fig. 3, where both vertices are of order $\epsilon^2$. The $N\Delta\gamma$
coupling stems from the counterterm lagrangian
\begin{equation}
{\cal L}_{\rm C.T.}= {i b_1\over 2 m} \bar\psi^\mu 
\left( g_{\mu\nu} + Y \gamma_\mu \gamma_\nu \right) \gamma_\rho \gamma_5
f_+^{\rho\nu} N \quad\rightarrow 
{\cal A}_{\Delta N}^{(2),\lambda}={i b_1 \over m} S_\nu f_+^{\nu\lambda},
\end{equation}
\begin{figure}
\begin{center}
\leavevmode
\hbox{%
\epsfxsize=10.0truecm
\epsfbox{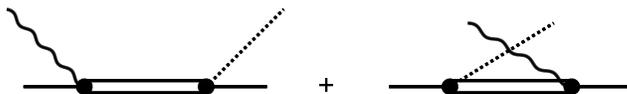} }
\end{center}
\caption{Contribution to the photoproduction amplitude at order $\epsilon^3$.
The dashed line denotes the neutral pion, the filled circles effective vertices
of order $\epsilon^2$, see text.}
\end{figure}
where I have indicated by the arrow its contribution to 
${\cal A}_{\Delta N}$ in (\ref{Seff}). The so called off-shell parameter 
Y does not contribute, as it is nonleading in $1/m$. The $N\Delta\pi$-vertex
in Fig. 3 arises from a $1/m$-correction contained in 
$-\left( \gamma_0 {\cal B}^\dagger_\Delta \gamma_0 {\cal C}_\Delta^{-1} 
{\cal B}_{\Delta N} \right)_\mu$ of (\ref{Atilde}) and is in proportion to 
$g_{\pi N\Delta}$. The net result is
\begin{equation}
E_{0+}={e \over {8\pi}}{4 b_1 g_{\pi N\Delta} \over {9F_\pi m^2}} 
{M_\pi^3 \over {M_\pi+\Delta}}.
\label{E0+new}
\end{equation}
In the chiral limit, this result scales like $M_\pi^3$, i.e. the corresponding 
photoproduction amplitude is of order $p^4$. There are {\it many} other terms of
this order, but (\ref{E0+new}) is the {\it only} term which is of order 
$\epsilon^3$. The LET of (\ref{LET}) is, of course, not violated by this term. 

Numerically, the term (\ref{E0+new}) turns out to 
be small. Using the coupling constant $b_1$ as determined
in \cite{BDM89}, $b_1=-2.30 \pm 0.35$, 
\footnote{In order to be consistent, the analysis of \cite{BDM89} should be 
redone in our formalism; work in this direction is under way.}
and employing the experimental result $g_{\pi N\Delta}=1.5\pm 0.2$,
the expansion of $E_{0+}$ reads
\begin{equation}
E_{0+}=-{e g_{\pi N} \over 8\pi m} \mu\left\{ 1-1.32+0.07 \right\} 
+O(\epsilon^4) = 0.80\times 10^{-3}/M_\pi
\end{equation}
where the second and third term in the brackets are the order $p^3$ and 
order $p^4$($\epsilon^3$) terms respectively. This result is far off the
number extracted \cite{Mei95} from the most recent experiment \cite{Fuchs95},
$E_{0+}=(-1.33\pm 0.08) 10^{-3}/M_\pi$.
It is therefore mandatory to calculate the next  
term in this expansion. Also, it will be most interesting to consider 
the energy dependence of $E_{0+}$ away from threshold. 
Work in this direction is under progress.

\section{Conclusions}

Heavy Baryon Chiral Perturbation Theory is a powerful tool to study the 
low energy interactions of pions, nucleons and photons. The spin 3/2 
delta resonances influence the effective low energy theory substantially,
due to the small mass difference $\Delta=m_N-m_\Delta$. The convergence 
of the chiral expansion can be studied by comparing HBChPT without explicit
spin 3/2 delta degrees of freedom with the low scale expansion sketched 
in section 3. This formalism treats all effects of the deltas, on- and 
off-shell, in a systematic manner and allows in particular to calculate
$1/m$-corrections in a consistent way. The many processes to which HBChPT
has already been applied are all well suited to be treated by the formalism 
presented here. Apart from phenomenological applications, further directions
of interest are the generalization to chiral SU(3), the complete one-loop
renormalization at the level of the functional integral as well as a better
understanding of the role of yet higher resonances.

\vspace{1cm}
\noindent
{\bf Acknowledgements}

\vspace{0.5cm}
\noindent
I would like to thank T. Hemmert and B. Holstein for the fruitful 
collaboration, M. Butler for useful comments at an early stage of this
work, and M.-T. Commault for the assistance in drawing the figures.

\end{document}